\documentclass{aa}
\usepackage{graphicx}
\usepackage{epsfig}
\usepackage{natbib}
\usepackage{amssymb}
\usepackage{amsmath}

\def\be{\begin{equation}}
\def\ee{\end{equation}}

\begin{document}

\title{\bf Constraints on variation in $\alpha$ and $m_e$ from WMAP 7-year data}

\author{Susana. J. Landau$^1$\thanks{member of  the Carrera del Investigador
    Cient\'{\i}fico y Tecnol\'ogico, CONICET.} \and Claudia G. Sc\'{o}ccola$^2$\thanks{Marie Curie fellow.}  } 

\institute{$^1$Departamento de F{\'\i}sica, FCEyN, Universidad de Buenos Aires,
  Ciudad Universitaria - Pab. 1, 1428 Buenos Aires, Argentina\\
 $^2$Max-Planck-Institut f\"ur Astrophysik, Karl-Schwarzschild Str. 1 Postfach 1317
D-85741 Garching, Germany.
\\
 \email{slandau@df.uba.ar,scoccola@mpa-garching.mpg.de}}

\date{Received; accepted}

\abstract{}{We update the constraints on the time variation
of the fine structure constant $\alpha$ and the electron mass $m_e$,
using the latest CMB data, including the 7-yr release of WMAP.}
{We made statistical analyses of the variation of each one of the
  constants and of their joint variation, together with the basic set
  of cosmological parameters. We used a modified version of CAMB and
  COSMOMC to account for these possible variations.}
{We present bounds on the variation of the constants for different
  data sets, and show how results depend on them. When using the
  latest CMB data plus the power spectrum from Sloan Digital Sky
  Survey LRG, we find that $\alpha / \alpha_0=0.986 \pm 0.007$ at
  1-$\sigma$ level, when the 6 basic cosmological parameters were
  fitted, and only variation in $\alpha$ was allowed. The constraints
  in the case of variation of both constants are $ \alpha / \alpha_0=
  0.986 \pm 0.009$ and $m_e / m_{e0} = 0.999 \pm 0.035$. In the case
  of only variation in $m_e$, the bound is $m_e / m_{e0}=0.964 \pm
  0.025$.  }
{}
 

\authorrunning{Landau \& Sc\'occola}

\titlerunning{ Constraints on variation in $\alpha$ and $m_e$ from WMAP 7-year data}

\maketitle

\section{Introduction}
\label{Intro}

The variation of fundamental constants over cosmological time scales
is a prediction of theories that attempt to unify the four
interactions in nature, like string derived field theories, related
brane-world theories and Kaluza-Klein theories (see \citet{Uzan03,Garcia-Berro07} and
references therein).  Many observational and experimental efforts have
been made to put constraints on such variations. Most of the reported
data are consistent with null variation of fundamental
constants. Although there have been recent claims for time variation
of the fine structure constant ($\alpha$) and of the proton to electron
mass ratio ($\mu =
\frac{m_p}{m_e}$) \citep{Murphy03b,Reinhold06}, independent analyses of
similar data give null results
\citep{Srianand04,King08,Thompson09,Malec10}.  On the other hand, a
recent analysis of ammonia spectra in the Milky Way suggests a spatial
variation of $\mu$ \citep{Molaro09,Levshakov10}.

Unifying theories predict variation of all coupling constants, being
all variations related in general to the rolling of a scalar
field. Therefore, the relationship between variations of coupling
constants depends on the unifying model. In this paper we adopt a
phenomenological approach and analyse the possible variation of
$\alpha$ and/or $m_e$ at the time of the formation of neutral hydrogen
without assuming any theoretical model. \citet{Nakashima10} have 
considered also the variation in the proton mass ($m_p$). This quantity
affects mainly the baryon mass density and the baryon number
density. Their results confirm the strong degeneracy with the baryon
density. Therefore, we will not consider the variation in $m_p$ in
this work.

Cosmic microwave background radiation (CMB) is one of the most
powerful tools to study the early universe and in particular, to put
bounds on possible variations in the fundamental constants between
early times and the present. Changing $\alpha$ or $m_e$ at
recombination affects the differential optical depth of the photons
due to Thompson scattering, changing therefore Thompson scattering
cross section and the ionization fraction.  The signatures on the CMB
angular power spectrum due to varying fundamental constants are similar to those
produced by changes in the cosmological parameters, i.e. changes in
the relative amplitudes of the Doppler peaks and a shift in their
positions. Moreover, an increment in $\alpha$ or $m_e$ decreases the
high-$\ell$ diffusion damping, which is due to the finite thickness of
the last-scattering surface, and thus, increases the power on very
small scales \citep{Turner,Hannestad99}.

Recent analysis of CMB data (earlier than the WMAP seven-year
release) including a possible variation in $\alpha$ have been
performed by \citet{Scoccola08,Scoccola09,Menegoni10,Nakashima10,Martins10},
and including a possible variation in $m_e$ have been performed by
\citet{Scoccola08,Scoccola09,Nakashima10}. 

In our previous works, we have also analyzed the dependence of the
updated recombination scenario (that includes the recombination of
helium, and was implemented in R{\sc ecfast} following \citet{wong08}) on
$\alpha$ and $m_e$, and show that these dependencies are not relevant
for WMAP data.

In this paper we adopt a phenomenological approach and analyse the
possible variation in $\alpha$ and/or $m_e$ without assuming any
theoretical model. We use WMAP seven-year release, together with other
recent CMB data. We also combine CMB data with other cosmological data
sets: i) the power spectrum of the Sloan Digital Sky Survery DR7 LRG,
ii) a recent constraint of the Hubble constant $H_0$ with data from
the Hubble Space Telescope.  In section \ref{statistics} we describe
the method and data sets we used in the statistical analysis. We
present and discuss our results in section \ref{discuss}.
We conclude in section \ref{conclusions}.

\section{Statistical Analysis}
\label{statistics}

We performed our statistical analysis by exploring the parameter space
with Monte Carlo Markov chains generated with the CosmoMC code
\citep{LB02} which uses the Boltzmann code CAMB \citep{LCL00} and
R{\sc ecfast} to compute the CMB power spectra. We modified them in
order to include the possible variation in $\alpha$ and $m_e$ at
recombination.

We use data from the WMAP 7-year temperature and
temperature-polarization power spectrum \citep{WMAP7}, and other CMB
experiments such as CBI \citep{CBI04}, ACBAR \citep{ACBAR02}, 
BOOMERANG \citep{BOOM05_polar,BOOM05_temp}, BICEP \citep{BICEP} and
QUAD \citep{QUAD}.  In order to reduce degeneracies of the
cosmological parameters, we combine the CMB data sets with other
cosmological data: i) the power spectrum of the Sloan Digital Sky
Survey LRG \citep{Reid09} and ii) the recent constraint on the Hubble
constant, $H_0= 74.2 \pm 3.6$ km s$^{-1}$ Mpc$^{-1}$, presented by
\citet{Riess09}. 
%

We have considered a spatially-flat
cosmological model with adiabatic density fluctuations, and the
following parameters:
\begin{equation}
P=\left(\Omega_b h^2, \Omega_{CDM} h^2, \Theta, \tau, \frac{\alpha}{\alpha_0}, \frac{m_e}{m_{e0}}, n_s, A_s\right) \nonumber \, \, 
\end{equation}
where $\Omega_b h^2$ is the baryon density and $\Omega_{CDM} h^2$ is
the dark matter density, both in units of the critical density;
$\Theta$ gives the ratio of the co-moving sound horizon at decoupling
to the angular diameter distance to the surface of last scattering (and is related to the Huble constant $H_0$);
$\tau$ is the reionization optical depth;
$n_s$ the scalar spectral index; and $A_s$ is the amplitude of the
density fluctuations.

We have performed statistical analyses using the data mentioned above
and considering variation of only one constant ($\alpha$ or $m_e$) and
variation of both constants. We present our results in the next
section.

\section{Results and Discussion}
\label{discuss}

 Results for the variation of the constants in the case when only one
 constant is allowed to vary are shown in Table~\ref{table-results1}
 and for the case when both are allowed to vary, are presented in
 Table~\ref{table-results2}. The obtained values are consistent with
 no variation of $\alpha$ or $m_e$ at recombination. The obtained
 errors are at the same percent level than those obtained by
 \citet{Scoccola08,Scoccola09,Menegoni10,Martins10} using WMAP-5 year
 release. The parameter space has higher dimension when both constants
 are allowed to vary. Therefore, limits on $\alpha$ and $m_e$ are more
 stringent in the case were only one constant is allowed to vary.
   Results for the cosmological parameters have similar values for all
   of the analyses. Therefore, we only report the values obtained in
   the case where both $\alpha$ and $m_e$ were allowed to vary and the
   data from CMB and the power spectrum of the SDSS DR7 were
   considered (see Table~\ref{pcosmo}). The mean values and errors for
   the cosmological parameters are in agreement within 1-$\sigma$ with
   those obtained by the WMAP collaboration \citep{WMAP7} with no
   variation of fundamental constants.

\begin{table}[!ht]
\begin{center}
\renewcommand{\arraystretch}{1.5}
\caption{Mean values and 1-$\sigma$ errors for the analysis with
  variation of only $\alpha$, and only $m_e$.}\label{table-results1}
\begin{tabular}{lcc}
\hline
\hline
Data set         & $\alpha/\alpha_0$ & $m_e/m_{e0}$  \\
\hline
all CMB          &  0.987$_{-0.009}^{+0.010}$       & 0.983$_{-0.066}^{+0.067}$  \\
all CMB + $H_0$    &  0.998$_{-0.007}^{+0.006}$       & 1.012$_{-0.018}^{+0.017}$  \\
all CMB + Sloan $P(k)$ &  0.986 $\pm$ 0.007  &   0.964 $\pm$ 0.025 \\
\hline
\end{tabular}
\end{center}
\end{table}

\begin{table}[!ht]
\begin{center}
\renewcommand{\arraystretch}{1.5}
\caption{Mean values and 1-$\sigma$ errors for the analysis with
 the joint  variation of $\alpha$ and $m_e$.}\label{table-results2}
\begin{tabular}{lcc}
\hline
\hline
Data set         & $\alpha/\alpha_0$ & $m_e/m_{e0}$   \\
\hline
all CMB               &   0.986 $\pm$ 0.010  &  1.015$_{-0.074}^{+0.075}$  \\
all CMB + $H_0$         &   0.986 $\pm$ 0.010  & 1.044 $\pm$ 0.029  \\
all CMB + Sloan $P(k)$  &   0.986 $\pm$ 0.009  & 0.999 $\pm$ 0.035   \\
\hline
\end{tabular}
\end{center}
\end{table}

\begin{table}
\caption{Mean values and 1$\sigma$ errors for the cosmological parameters using all CMB data and the SDSS DR7 power spectrum. $H_0$ is in units of km s$^{-1}$ Mpc$^{-1}$.} \label{pcosmo}
\renewcommand{\arraystretch}{1.5}
\begin{center}
\begin{tabular}{lc}
\hline
\hline
  parameter &  all CMB + SDSS \\
\hline
 $\Omega_b h^2$ & 0.02195$_{-0.00068}^{+0.00067}$    \\
 $\Omega_{CDM} h^2$ & 0.1070$_{-0.0065}^{+0.0065}$    \\
$\tau$ & 0.087$_{-0.007}^{+0.006}$    \\
$n_s$ & 0.971$_{-0.013}^{+0.013}$ \\
$A_s$ & 3.097$_{-0.036}^{+0.035}$ \\
$H_0$ &  64.3$_{-4.4}^{+4.3}$    \\
\hline
\end{tabular}
\end{center}
\end{table}

\begin{figure}[!ht]
\caption{68\% and 95\% c.l. constraints for $\alpha/\alpha_0$ versus $H_0$, for the analysis of the variation of $\alpha$ alone. Results  from different data sets.  }\label{fig:alfa-H0}
\begin{center}
\includegraphics[scale=.8,angle=0]{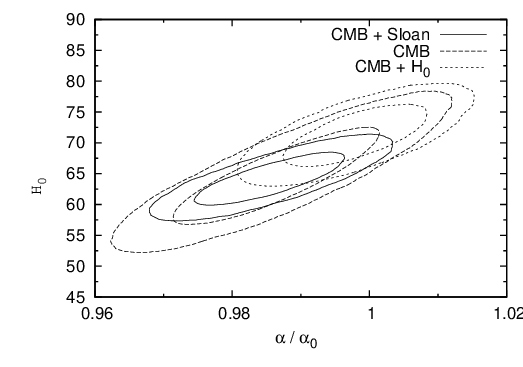}
\end{center}
\end{figure}

In Fig.~\ref{fig:alfa-H0} we show the 68\% and 95\% c.l. constraints
for $\alpha/\alpha_0$ versus $H_0$, for the analysis of the variation
of $\alpha$ alone. The results correspond to different data sets: all
the CMB data alone; all the CMB data plus the $H_0$ prior taken from
\citet{Riess09}; and all the CMB data plus the power spectrum from
Sloan Digital Sky Survery DR7 LRG \citep{Reid09}. The large degeneracy
between $\alpha/\alpha_0$ and $H_0$ from CMB data is reduced when
another data set is added. However, since the value of $H_0$ obtained
from the extra data sets are different, the obtained constraint on
$\alpha/\alpha_0$ depends strongly on the data chosen for the
analysis. Nevertheless, the results are consistent within
1-$\sigma$.

\begin{figure}[!ht]
\caption{68\% and 95\% c.l. constraints for $\alpha/\alpha_0$ versus $\tau$, for the analysis of the variation of $\alpha$ alone. Results  from different data sets.  }
\begin{center}
\includegraphics[scale=.8,angle=0]{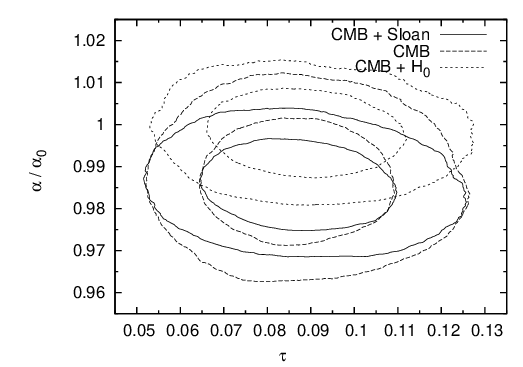}
\end{center}
\label{fig:alfa-tau}
\end{figure}

\begin{figure}[!ht]
\caption{68\% and 95\% c.l. constraints for $\alpha/\alpha_0$ versus $\Omega_b h^2 $, for the analysis of the variation of $\alpha$ alone. Results  from different data sets.  }\label{fig:alfa-ob}
\begin{center}
\includegraphics[scale=.8,angle=0]{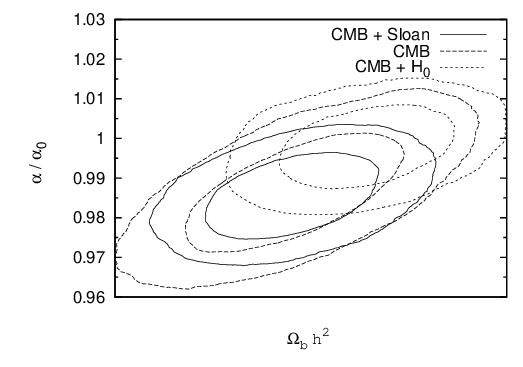}
\end{center}
\end{figure}

In Fig.~\ref{fig:alfa-tau} we present the constraints for
$\alpha/\alpha_0$ versus $\tau$ and in Fig.~\ref{fig:alfa-ob} we
present the constraints for $\alpha/\alpha_0$ versus $\Omega_b h^2$. There
are degeneracies among these parameters. The contours change because
of the different mean value of $\alpha/\alpha_0$ obtained with
different data sets.

\begin{figure}[!ht]
\caption{68\% and 95\% c.l. constraints for $m_e/m_{e0}$ versus $H_0$, for the analysis of the variation of $m_e$ alone. Results  from different data sets.  }\label{fig:emasa-H0}
\begin{center}
\includegraphics[scale=.8,angle=0]{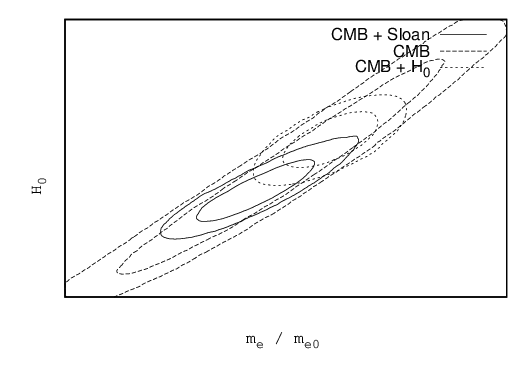}
\end{center}
\end{figure}

In Fig.~\ref{fig:emasa-H0} we present the result for the case where
only $m_e$ was allowed to vary.  The degeneracy between $m_e/m_{e0}$
and $H_0$ is larger than between $\alpha/\alpha_0$ and $H_0$, making
impossible to find reliable constraints using CMB data alone.  When
another data set is added, the bounds result tighter, but the mean
value for $m_e/m_{e0}$ depends strongly on which data set was
added. Results are marginally consistent at 1-$\sigma$.

\begin{figure}[!ht]
\caption{68\% and 95\% c.l. constraints for $m_e/m_{e0}$ versus $\tau$, for the analysis of the variation of $m_e$ alone. Results  from different data sets.} \label{fig:emasa-tau}
\begin{center}
\includegraphics[scale=.8,angle=0]{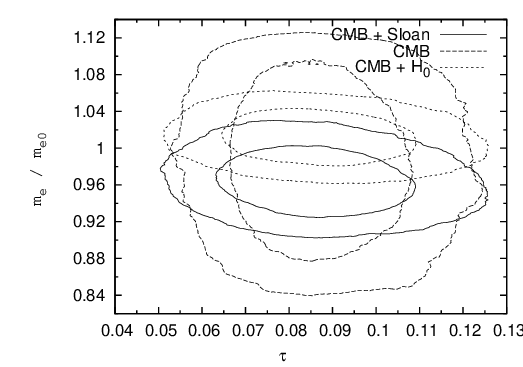}
\end{center}
\end{figure}

\begin{figure}[!ht]
\caption{68\% and 95\% c.l. constraints for $m_e/m_{e0}$ versus $\Omega_b h^2$, for the analysis of the variation of $m_e$ alone. Results  from different data sets.  } \label{fig:emasa-ob}
\begin{center}
\includegraphics[scale=.8,angle=0]{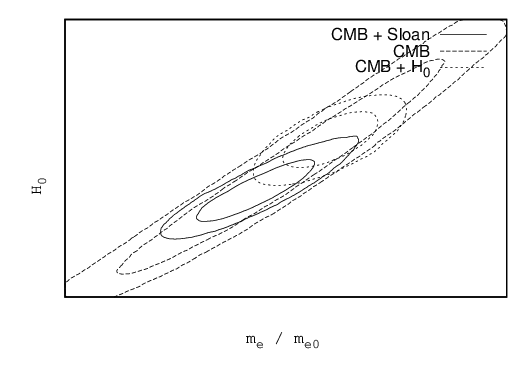}
\end{center}
\end{figure}

The constraints on $m_e/m_{e0}$ versus $\tau$ are shown in
Fig.~\ref{fig:emasa-tau}, and on $m_e/m_{e0}$ versus $\Omega_b  h^2$ are
shown in Fig.~\ref{fig:emasa-ob}. In both cases, the results depend on
the data set added to CMB data in the statistical analysis.

\begin{figure}[!ht]
\caption{68\% and 95\% c.l. constraints for the joint variation of $\alpha$ and $m_e$ from different data sets.  }\label{fig:alfa-emasa}
\begin{center}
\includegraphics[scale=.8,angle=0]{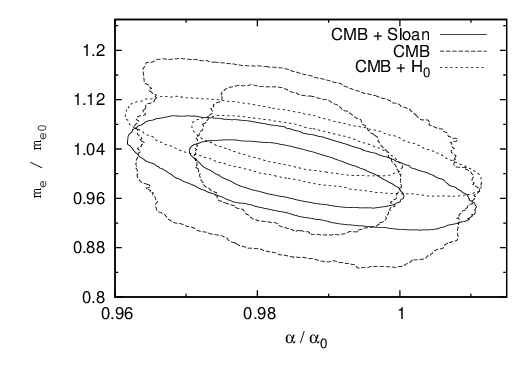}
\end{center}
\end{figure}

In Fig.~\ref{fig:alfa-emasa} we show the posterior distribution for
$\alpha/\alpha_0$ and $m_e/(m_e)_0$, for the case of joint variation
of these quantities, marginalized over the cosmological
parameters. The results correspond to different data sets. The
difference in the contours is mainly due to the large degeneracy of
$m_e$ and $H_0$, and the different $H_0$ values derived from the Sloan
power spectrum and from the $H_0$ prior.  We see that the mean value
of $m_e$ is more affected than the mean value of $\alpha$. These
results can also be seen in Table~\ref{table-results2}.

 A variation of $\alpha$ or $m_e$ affects the recombination scenario
 (see \citet{Scoccola09} for example).  As a consequence, the angular
 diameter distance at recombination is modified if any of these
 constants varies. This results in a change in the Doppler peak
 positions and heights (see \citet{Turner} for example). This explains
 the degeneracy between $\alpha$ and $m_e$ shown in
 Fig.~\ref{fig:alfa-emasa} and confirmed by the correlation
 coefficient. On the other hand, the degeneracy between $\alpha$ or
 $m_e$ with the baryon mass density or the Hubble constant can be
 explained since these effects are similar to a change in the
 cosmological parameters.  A variation in $\alpha$ and/or $m_e$
   at recombination, affects mainly the binding energy of
   hydrogen. This quantity is proportional to $m_e \alpha^2$. When
   only one constant is allowed to vary, its influence on the
   parameter estimation is similar, regardless of the constant.  However, when a joint variation
   analysis is performed, the results are different for $\alpha$ and
   $m_e$, due to the power with which they enter the hydrogen binding
   energy. In particular, in Fig.~\ref{fig:alfa-emasa}, we note that
   the bounds on $\alpha$ are not affected when including additional
   data sets to the CMB data. This is due to the fact that $\alpha$ is
   no longer correlated with $H_0$, as it was shown previously (see,
   for example, \citet{landau08}).

\section{Conclusions}
\label{conclusions}

In this paper we have updated the constraints on the time variation of
the fine structure constant $\alpha$ and the electron mass $m_e$
during recombination epoch, using the latest CMB data, including the
7-yr release of WMAP. We perform several statistical analyses adding
two different data sets; the $H_0$ prior taken from \citet{Riess09};
and the power spectrum from Sloan Digital Sky Survery DR7 LRG
\citep{Reid09}. The bounds on the variation of the constants are
tighter than previous results because of the higher precision of the
new data used in this work.

Our results show no variation of the constants at recombination time.
We also emphasize that the constraints depend strongly on which data
set we choose in the analysis, due to the large degeneracy between
$\alpha$ or $m_e$ and $H_0$. Yet, the results are consistent
within 1-$\sigma$.

\section*{{\bf Acknowledgements}}
 The research leading to these results has received funding from the
European Community's Seventh Framework Programme ([FP7/2007-2013]
under grant agreement n$^\circ$ 237739), and  PICT 2007-02184 from
Agencia Nacional de Promoci\'on Cient{\'\i}fica y Tecnol\'ogica, Argentina.
The authors would like to thank H.C. Chiang for help with compiling
the BICEP dataset in COSMOMC.

\newpage

\bibliography{14125}
\bibliographystyle{aa}
\end{document}